\documentclass[journal]{IEEEtran}
\pdfoutput=1
\usepackage[T1]{fontenc} 
\usepackage{amsmath}
\usepackage{amssymb}
\usepackage{bm}
\usepackage{cases}
\usepackage{graphicx}
\makeatletter
\newcommand{\removelatexerror}{\let\@latex@error\@gobble}
\makeatother

\usepackage[ruled,norelsize]{algorithm2e}

\usepackage{epsfig}
\usepackage{epstopdf}
\ifCLASSINFOpdf

\else

\fi
\begin{document}
%
\title{Interference Management and Power Allocation for NOMA Visible Light Communications Network}

\IEEEoverridecommandlockouts
\author{\IEEEauthorblockN{Xiaoke Zhang, Qian Gao, Chen Gong and Zhengyuan Xu}
\thanks{This article has been submitted to IEEE Communication Letters for publication on July 27, 2016.}
\thanks{This work was supported by National Key Basic Research Program of China (Grant No. 2013CB329201), National Natural Science Foundation of China (Grant No. 61501420), Shenzhen Peacock Plan (No. 1108170036003286), and the Fundamental Research Funds for the Central Universities (WK3500000001 and WK3500000003).}
\thanks{The authors are with Key Laboratory of Wireless-Optical Communications, Chinese Academy of Sciences, University of Science and Technology of China, Hefei, China. Z. Xu is also with Shenzhen Graduate School, Tsinghua University, Shenzhen, China. Email: zxiaoke@mail.ustc.edu.cn, \{qgao, cgong821, xuzy\}@ustc.edu.cn.}


}

\maketitle

\begin{abstract}
To design an efficient interference management and multiple access scheme for visible light communication (VLC) network, this letter leverages the non-orthogonal multiple access (NOMA), which has received significant attention in the $5^{th}$ generation wireless communication. With the residual interference from the successive interference cancellation in NOMA taken into account, we optimize the power allocation for NOMA VLC network to improve the achievable user rate under user quality of service (QoS) constraint. The performance of the proposed approaches is evaluated by the numerical results.
\end{abstract}
\IEEEpeerreviewmaketitle


\section{Introduction}
With a rapid growth in portable information terminals, the demand for high rate wireless data communication in local area networks keeps increasing. Although radio frequency has been widely commercialized for communication purposes because of the wide area coverage and little interference in frequency band \cite{karunatilaka2015led}, the limited spectrum cannot well accommodate the increasing communication rate requirement. Visible light communication (VLC) has received significant interest due to its advantages in unlicensed spectrum, natural confidentiality, convenient deployment, low energy consumption, etc. VLC network is also considered to be an environmental-friendly solution for smart home networking.

Considering the realistic network application, the VLC network shall support multi-user access. The main obstacle for efficient multiple access is the severe interference among the links and a few terminal access approaches have been proposed. A graph theory based scheduling method is proposed in \cite{tao2015scheduling} where the users are selected to access the network according to proportional fairness priority factor. The ``neighbouring'' users of an active user are muted to avoid excessive interference. Furthermore, \cite{7217841} proposes a user-centric (UC) cluster formation technique employing vectored transmission (VT) to allow each multiple-access-point cell serving multiple users simultaneously. Irregular-shape elastic cell formations are supported which are dynamically constructed and adjusted based on traffic requirements.

On the other hand, non-orthogonal multiple access (NOMA) has been recently suggested as a promising solution in the $5^{th}$ generation (5G) wireless networks. Multiple users are multiplexed in the power domain on the transmitter side by superposition coding and multi-user signal separation is accomplished on the receiver side by successive interference cancellation (SIC) at the receiver side. NOMA is demonstrated to outperform the orthogonal multiple access in terms of the ergodic sum rate and the user outage probability \cite{ding2014performance}. With fixed power allocation, the performance of NOMA for a single VLC access point (VAP) is investigated in \cite{yin2015performance}. The channel dependent gain ratio power allocation is proposed in \cite{marshoud2015non} to ensure efficiency and fairness compared to static power allocation approach.

The contribution of this letter is two-fold. Firstly, we analyze the realistic NOMA-VLC network scheme and propose the corresponding interference management strategy based on the user locations. Secondly, we propose two types of quality of service (QoS) guaranteed power allocation for NOMA based on the sum user rate criterion or the max-min user rate criterion. The residual interference during the SIC in NOMA is taken into account in the formulated optimization problem.


\section{NOMA VLC Network}

\subsection{Multiple Access and Interference Management Scheme}

Consider the downlink communication scenario where multiple VLC access points (VAPs) are placed in the ceiling and multiple mobile users are randomly distributed within the circular coverage area underneath. Distinctive from traditional radio frequency communications, the cell in VLC is significantly smaller for the sharp VLC signal attenuation and restricted user FOV, which brings a big challenge for interference management among adjacent co-frequency cells in network application. NOMA has the congenital advantage of supporting multiple access which brings great convenience to mobile network construction.

\begin{figure}[!t]
\centering
\includegraphics[width=3.0in]{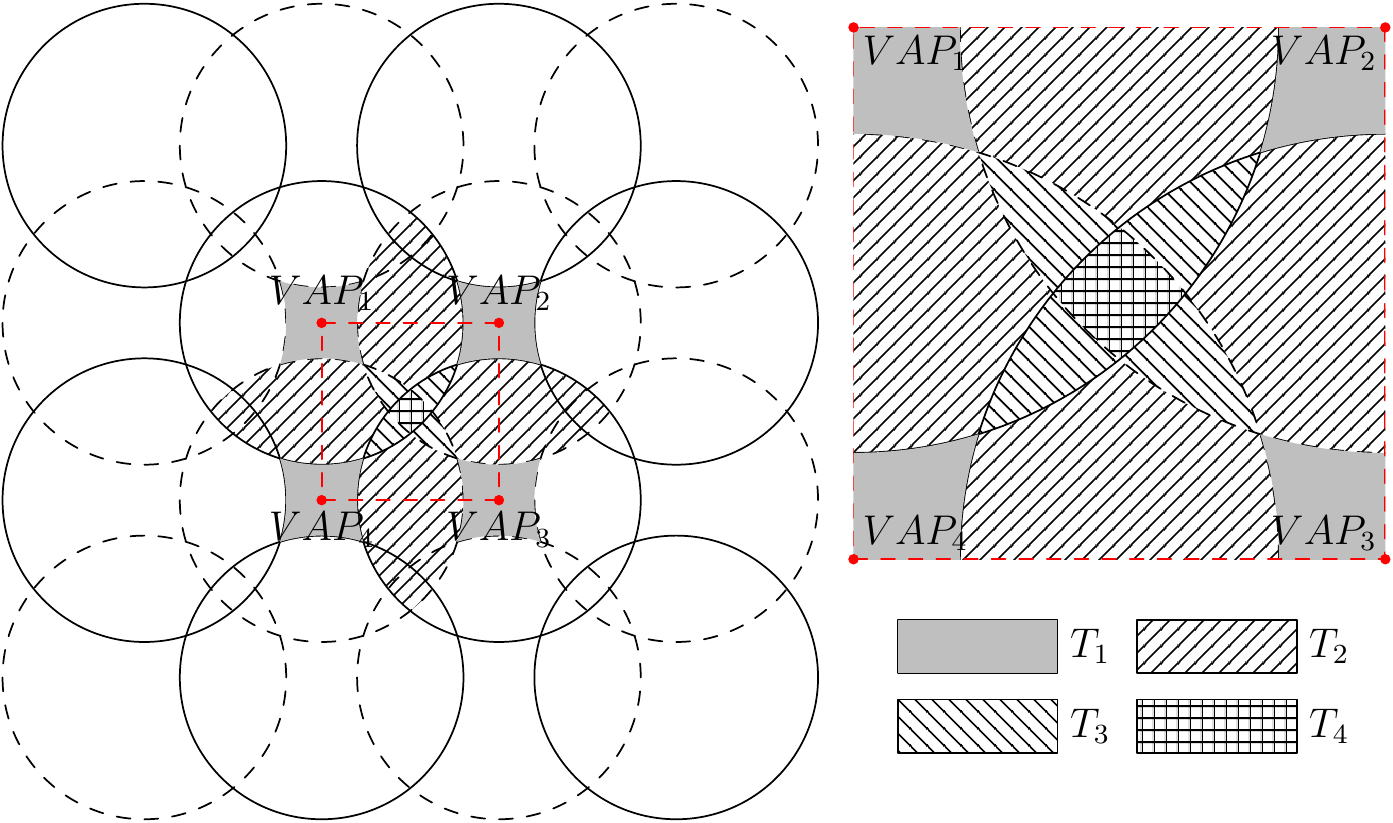}
\caption{NOMA VLC network model}
\label{NOMA_VLC_network}
\end{figure}

A typical NOMA VLC network is shown in Figure \ref{NOMA_VLC_network}. The red dots denote the network VAPs in the network. The coverage area consists of four types. Area type $L_i,i\in\{1,2,3,4\}$ represents that the corresponding hatched area can receive LOS signal of $i$ VAPs. The neighboring cells are distinguished by solid line and dashed line. The cell size is determined by the user FOV, vertical distance, attenuation coefficient and so on.
The interference is influenced by the following factors:
\begin{enumerate}
  \item \textit{User FOV: } If the user FOV is sufficiently small, there is only at most one VAP in the FOV which will certainly eliminate the inter-cell interference and the remaining intra-cell interference can be managed by the multi-user access policy. However, this would potentially cause coverage holes where the communication may fail because of no LOS optical signal. Thus, the FOV needs to be adjusted to an appropriate value.
  \item \textit{User Distribution: } It is easy to see that the inter-cell interference can be avoided without co-frequency VAPs working. Certainly, this condition depends on the time-varying user distribution.
  \item \textit{Frequency Reuse (FR) Factor: } Although $FR=1$ can achieve the highest spectrum efficiency, users located in areas $L_2,L_3,L_4$ in Figure \ref{NOMA_VLC_network} may suffer from severe interference. To solve such an issue, sophisticated interference cancellation technique or user scheduling is needed in order to accomplish the communication.  This situation would be significantly improved if $FR=2$ since only users in area type $L_4$ cannot avoid co-frequency interference.
\end{enumerate}

To balance the interference management requirements and the frequency efficiency, we adopt $FR=2$ in this letter. Assume that the available bandwidth for each cell is $B$ and uniform transmission power supply is adopted among the VAPs. Based on the user positions, the multiple access scheme for NOMA VLC network is designed as follows. The users in $L_1$ and $L_3$ are assigned to the VAP without interference. As the probability of users existing in $L_4$ is reasonably small, the users in $L_4$ are allocated with special bandwidth $B_i$ to avoid the interference. Since a user in $L_2$ can select either VAP$_1$ or VAP$_2$ for access without interference, they are scheduled for load balancing among the VAPs and $1/d(VAP_i)$ may be regarded as the priority, where $d(VAP_i)$ denotes the number of users connected to VAP$_i$.

\subsection{System Model}

The power of the reflected signal is usually much weaker than that of the line-of-sight (LOS) signal and thus can be neglected. The VAP is placed at height $L_k$ above the users. The $k$-th user, denoted as $U_k$, is located on a polar coordinate plane at the distance $d_k$ from the LED, and the LED irradiance angle and the PD incidence angle are given by $\phi_k$ and $\psi_k$, respectively. According to the Lambertian emission model, the channel gain of the optical link between the VAP and the $k$-th user, denoted as $h_k$, is given by
\begin{equation}
    h_k=\frac{A(m+1)}{2 \pi d_k^2}\cos^m(\phi_k)T(\psi_k)g(\psi_k)\cos(\psi_k),
\end{equation}
where $A$ denotes the detection area of the PD, $T(\psi_k)$ represents the gain of the optical filter, $m$ is the order of Lambertian emission relying on the transmitter semiangle $\Phi_{1/2}$ by $m=-\ln2/\ln(\cos\Phi_{1/2})$. The gain of nonimaging concentrator $g(\psi_k)$ with refractive index $n$ is given by ${n^2}/{\sin^2(\Psi_c)}$ if $0 \leq \psi_k \leq \Psi_c$ and $0$ otherwise, where $\Psi_c$ is the concentrator field of view (FOV) semiangle \cite{karunatilaka2015led}.

Without loss of generality, assume that the $K$ users in a particular cell are sorted based on the link gain as $h_1 \leq h_2 \leq \cdots \leq h_K$. Let $s_k$ denote the message that the VAP delivers to $U_k$. According to the NOMA protocol, $\{s_k, k=1,2,\ldots,K\}$ is superposed and transmitted by the VAP as
\begin{equation}
  x=\sum_{i=1}^{K}a_i\sqrt{P_{elec}}s_i+I_{DC},
\end{equation}
where $I_{DC}$ is the DC bias added to ensure the positive instantaneous intensity and $a_i$ is the power allocation coefficient for $i$-th user. Therefore, the observation at the $k$-th user is given by
\begin{equation}
  y_k=h_k\sum_{i=1}^{K}a_i\sqrt{P_{elec}}s_i+n_k,
\end{equation}
where $n_k$ denotes the additive real-valued Gaussian noise with zero mean and variance $\sigma_k^2$ including the shot noise and the thermal noise.

In NOMA, users with low channel gain will be allocated more power, i.e. $a_1 \geq a_2 \geq \cdots \geq a_K$. At the receiver side, the user would perform successive interference cancellation \cite{ding2014performance}. Note that the SIC requires highly accurate channel and signal estimation otherwise the non-negligible residual interference remains. Although optical wireless channel is practically stationary, the channel estimation error can still exist due to the feedback delay and user mobility. As a result, to ensure each user's QoS requirements, the achievable rate for the $k$-th user is given by
\begin{equation}
    \label{noma_condition_imperfect}
    \tilde{R}_{k \rightarrow j}=
    \begin{cases}
        \log_2\left(1 + \frac{\left( h_k a_j \right)^2}
        {\sum_{i=j+1}^{K}\left( h_k a_i \right)^2 +
        \varepsilon \sum_{i=1}^{j-1}\left( h_k a_i \right)^2+
        1/\rho}
        \right) \geq T_{j},\\
        \qquad j \leq k,j \neq K;\\
        \log_2 \left( 1+\frac{\left( h_K a_j \right)^2}
        {\varepsilon \sum_{i=1}^{K-1}\left( h_k a_i \right)^2+1/\rho}\right)
        \geq T_{j} , \quad j=k=K;\\
    \end{cases}
\end{equation}
where $\rho=\gamma^2 \cdot TSNR$, $TSNR=P_{elec}/N_0B$ denotes the transmitted signal-to-noise ratio (TSNR) \cite{ding2014performance}\cite{yin2015performance}, $\gamma$ denotes the photoelectric conversion efficiency, $T_{j}$ denotes the targeted data rate for successful message detection at the $j$-th user, $\tilde{R}_{k \rightarrow j}$ denotes the achievable data rate for the $k$-th user to detect the $j$-th user's message and $\varepsilon$ represents the residual interference coefficient \cite{andrews2003optimum}.

Note that if $k_1 \geq k_2 \geq i$, then $\tilde{R}_{k_1 \rightarrow i} \geq \tilde{R}_{k_2 \rightarrow i}$. Therefore, Eq. \eqref{noma_condition_imperfect} can be simplified as
\begin{equation}
    \label{sim_condition}
    \tilde{R}_{k} \geq T_{k}, \quad k=1,2,\cdots,K,
\end{equation}
where $\tilde{R}_{k} \triangleq \tilde{R}_{k \rightarrow k}$. It can be seen from Eq. \eqref{noma_condition_imperfect} and \eqref{sim_condition} that the power allocation parameters $\{a_1,a_2,\ldots,a_K\}$ jointly determine each user's achievable rate and thus may nontrivially affect the corresponding modulation and coding scheme for data transmission of each user.
As the power allocation plays a key rule in NOMA, we investigate the QoS-guaranteed power allocation strategies for NOMA in Section \ref{sec:power_allocation}.

\section{QoS-guaranteed NOMA power allocation}
\label{sec:power_allocation}

\subsection{QoS-guaranteed Max-Sum Rate Criterion}

NOMA can support a flexible management of user rate and provide an efficient way to ensure fairness by adjusting power allocation coefficients. We aim to maximize the sum of user achievable rate by optimizing the power allocation while satisfying the basic QoS requirements. The corresponding optimization problem is formulated as follows,
\begin{align}
    \label{sum_rate_origin}
    \begin{aligned}
    \text{maximize} \quad &\sum_{k=1}^{K} \tilde{R}_k,\\
    \text{s.t.} \quad &
    \begin{cases}
        &\tilde{R}_k \geq T_k,\quad k=1,2,\ldots,K;\\
        &\sum_{k=1}^{K} a_k^2=1;\\
        &a_k \geq 0, \quad k=1,2,\ldots,K.
    \end{cases}.
    \end{aligned}
\end{align}
Notice that $\tilde{R}_k$ is a function of $\{a_k,a_{k+1},\ldots,a_K\}$, which leads to the objective parameters coupling in the original optimization problem and brings difficulty to analysis. Perform variable substitution as $s_k  \triangleq \sum_{i=k}^{K} a_k^2, \; k=1,2,\ldots,K$ and Eq. \eqref{noma_condition_imperfect} can be equivalently expressed as
\begin{align}
    \label{imperfect_rate_equ}
    \tilde{R}_k=\begin{cases}
    \begin{aligned}
        &\log_2\left(
            \frac{(1-\varepsilon)s_k+m_k}
            {s_{k+1}-\varepsilon s_k+m_k}
        \right) \triangleq G_k(s_k,s_{k+1}),\\
        & \qquad k=1,2,\ldots,K-1;\\
        & \log_2\left(
            \frac{(1-\varepsilon)s_K+m_K}
            {-\varepsilon s_K+m_K}
        \right) \triangleq G_K(s_K),\\
        & \qquad k=K;
    \end{aligned}
    \end{cases},
\end{align}
where $m_k=\varepsilon+1/\left(\rho h_k^2\right),\; k\in\{1,2,\ldots,K\}$. Then the achievable sum rate of users is denoted as $\tilde{R}_{total}$ and transformed into separable form as $\tilde{R}_{total}=\sum_{k=1}^{K-1}G_k(s_k,s_{k+1})+G_K(s_K)$.

The constraints in problem \eqref{sum_rate_origin} form the feasible region $\mathfrak{D}$. Clearly, these constraints are linear and are hence convex. However, the objective function is not convex which is difficult to solve directly using standard optimization solvers. We develop a gradient projection (GP) algorithm \cite{boyd2004convex} which includes a gradient descending process and a projection process.

Let $\bm{s}$ denote the variable $\bm{s}=\left(s_2,\cdots,s_K\right)$. The gradient descending process iteratively takes steps in the direction of the gradient of the objective function at a given position yielding
\begin{equation}
    \widetilde{\bm{s}}^{(i+1)}=\bm{s}^{(i)}+\lambda_{i} \cdot \frac{\partial \tilde{R}_{total}(\bm{s})}
    {\partial \bm{s}}\Big|_{\bm{s}=\bm{s}^{(i)}},
\end{equation}
where the superscript $(\cdot)^{(i)}$ denotes the iteration time, $\widetilde{\bm{s}}$ denotes the variable with step added and $\lambda_{i}$ denotes the step size which can be chosen by backtracking line search \cite{boyd2004convex}. When the variable $\widetilde{\bm{s}}^{(i+1)}$ steps out of $\mathfrak{D}$, it is mapped into $\mathfrak{D}$ by finding the nearest feasible point in $\mathfrak{D}$. The corresponding projection process is described as a convex optimization problem whose solution can be efficiently obtained utilizing the standard solver such as CVX \cite{grant2008cvx} embedded with MATLAB$^{\textregistered}$.


\subsection{QoS-guaranteed Max-min Rate Criterion}

Distinct from the fairness criterion investigated in \cite{timotheou2015fairness} where the minimum of user achievable rate is maximized, we consider additional QoS requirement of each user and take the residual interference into account. We propose the QoS-guaranteed max-min rate criterion with the associated optimization problem formulated as follows
\begin{align}
    \label{maxmin}
    \max_{\{a_1,a_2,\ldots,a_K\}} & \min_{i \in \{1,2,\ldots,K\}} \quad \tilde{R}_i,\\
    \text{s.t.} \quad & \begin{cases}
    \begin{aligned}
     &\tilde{R}_k \geq T_k,\quad k=1,2,\ldots,K;\\
     &\sum_{k=1}^{K} a_k^2=1;\\
     &a_k \geq 0, \quad k=1,2,\ldots,K;
    \end{aligned}
    \end{cases}
\end{align}
With the additional QoS constraints, this problem can still be solved following the searching algorithm in \cite{timotheou2015fairness} which is time-consuming to obtain a solution with desired accuracy. Instead, we adopt GP algorithm to dynamically adjust the power allocation parameters in a short interval. To overcome the non-differentiability of objective function in \eqref{maxmin}, we adopt the following approximation \cite{drost2014constellation},
\begin{align}
    \label{minimum_app}
    \min_{i \in \{1,2,\ldots,K\}} R_i \simeq \lim_{\beta\rightarrow + \infty} \frac{1}{-\beta}
    \ln\left( \sum_{i=1}^{K} \exp(-\beta R_i) \right).
\end{align}

In summary, to fulfill the basic QoS requirements of users, we propose two QoS-guaranteed user rate optimization criteria, which maximize the sum user rate and the minimum user rate. For both criteria, the residual interference from SIC is taken into consideration. The maximum sum user rate criterion is proposed to improve the overall network throughput and the max-min rate criterion aims to balance the fairness and the user rate demand.


\begin{table}
    \caption{Simulation parameters}
    \label{Sim_para_table}
    \centering
    \begin{tabular}{c|c}
    \hline
    Parameter name, notation & Value\\
    \\ \hline \hline
    VAP height, $H$ & $3$ m \\
    User height, $z$ & $0.85$ m \\
    Semi-angle at half power, $\phi_{1/2}$ & $60^{\circ}$ \\
    PD detection area, $A$ & $1$ $\text{cm}^2$ \\
    PD responsivity, $\gamma$ & $0.4$ A/W \\
    PD FoV, $\Psi_{fov}$ & $32^{\circ}$ \\
    Optical filter gain, $T(\psi)$ & $1$ \\
    Refractive index, $n$ & $1.5$ \\
    Transmitted SNR, $TSNR$ & $65 \sim 85\text{dB}$\\
    \hline
    \end{tabular}
\end{table}


\section{Numerical Results}

In this section, we evaluate the performance of our proposed QoS-guaranteed system with max-sum rate and max-min rate criteria parameterized as in Table \ref{Sim_para_table}.

\begin{figure}[!t]
\centering
\includegraphics[width=3.0in]{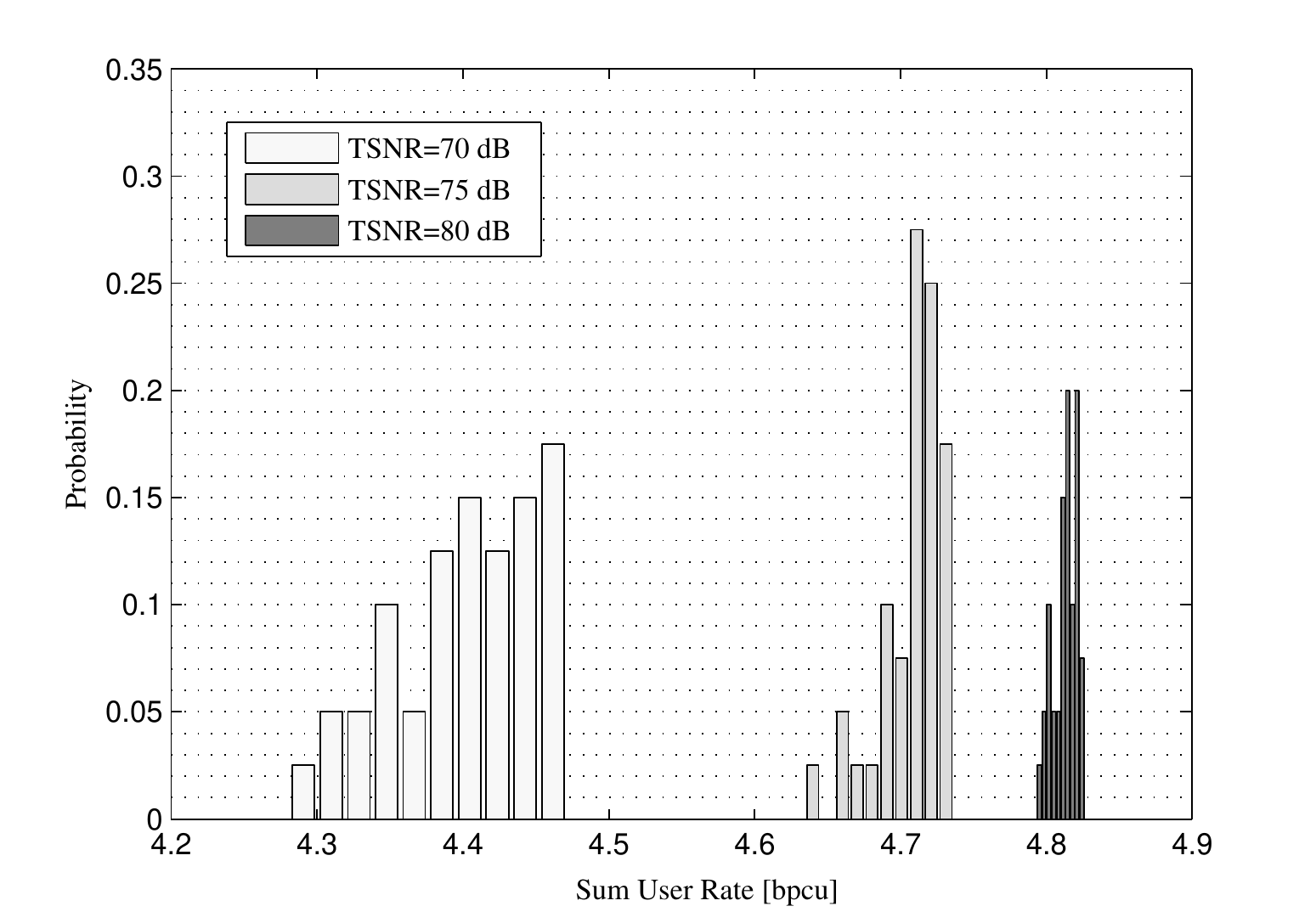}
\caption{The distribution of the sum user rate for random deployed users when $K=3$, $\varepsilon=0.06$, $\bm{T}=[0.6,0.6,0.6]$.}
\label{H_random_test}
\end{figure}

\begin{figure}[!t]
\centering
\includegraphics[width=3.0in]{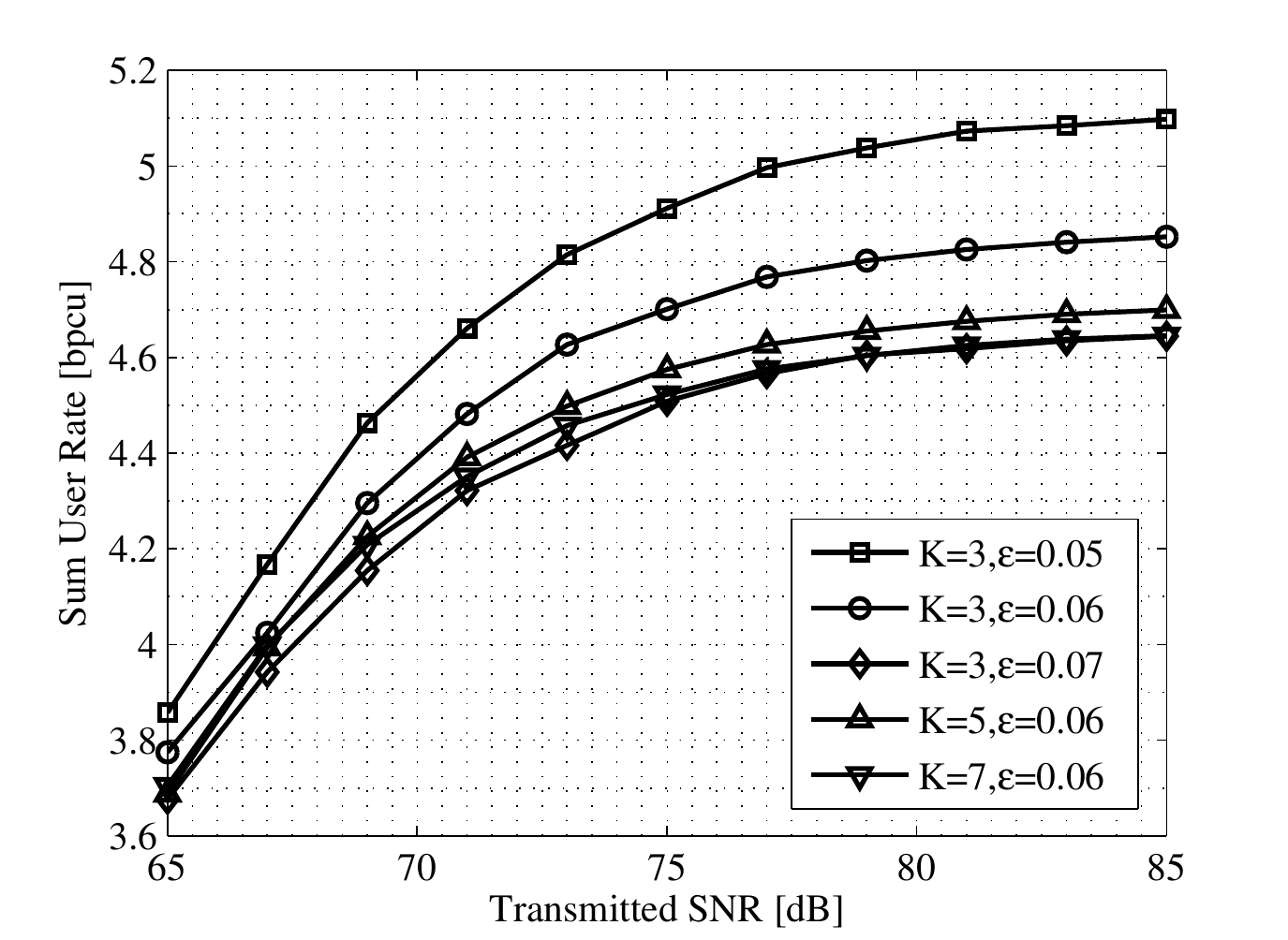}
\caption{The sum user rate comparison under different user number $K$ and residual interference coefficient $\varepsilon$ values when $\bm{T}=[0.6,0.6,0.6]$.}
\label{sum_rate_SNR}
\end{figure}

Assume that the $K$ users simulated are uniformly distributed in a VAP coverage area. Figure \ref{H_random_test} compares the statistical distribution of the sum user rate for various $TSNR$ values for $K=3$ and $\varepsilon=0.06$. As the $TSNR$ increases, the sum rate is improved due to less noise while the variance is reduced since the noise item $1/(\rho h^2)$ has less impact on $m=\varepsilon+1/\left(\rho h^2\right)$ compared to the residual interference. This implies that the user distribution has a larger influence on the sum user rate when lower $TSNR$ exists. The number of users $K$ and the residual interference coefficient $\varepsilon$ are another two factors that affect the NOMA performance. As Figure \ref{sum_rate_SNR} shows, the max-sum rate increases as $K$ decreases, which complies with the smaller number of users in VAP coverage area. Moreover, a slight increase in the residual interference significantly degrades the NOMA performance, which emphasizes the importance of the channel estimation accuracy.

\begin{figure}[!t]
\centering
\includegraphics[width=3.0in]{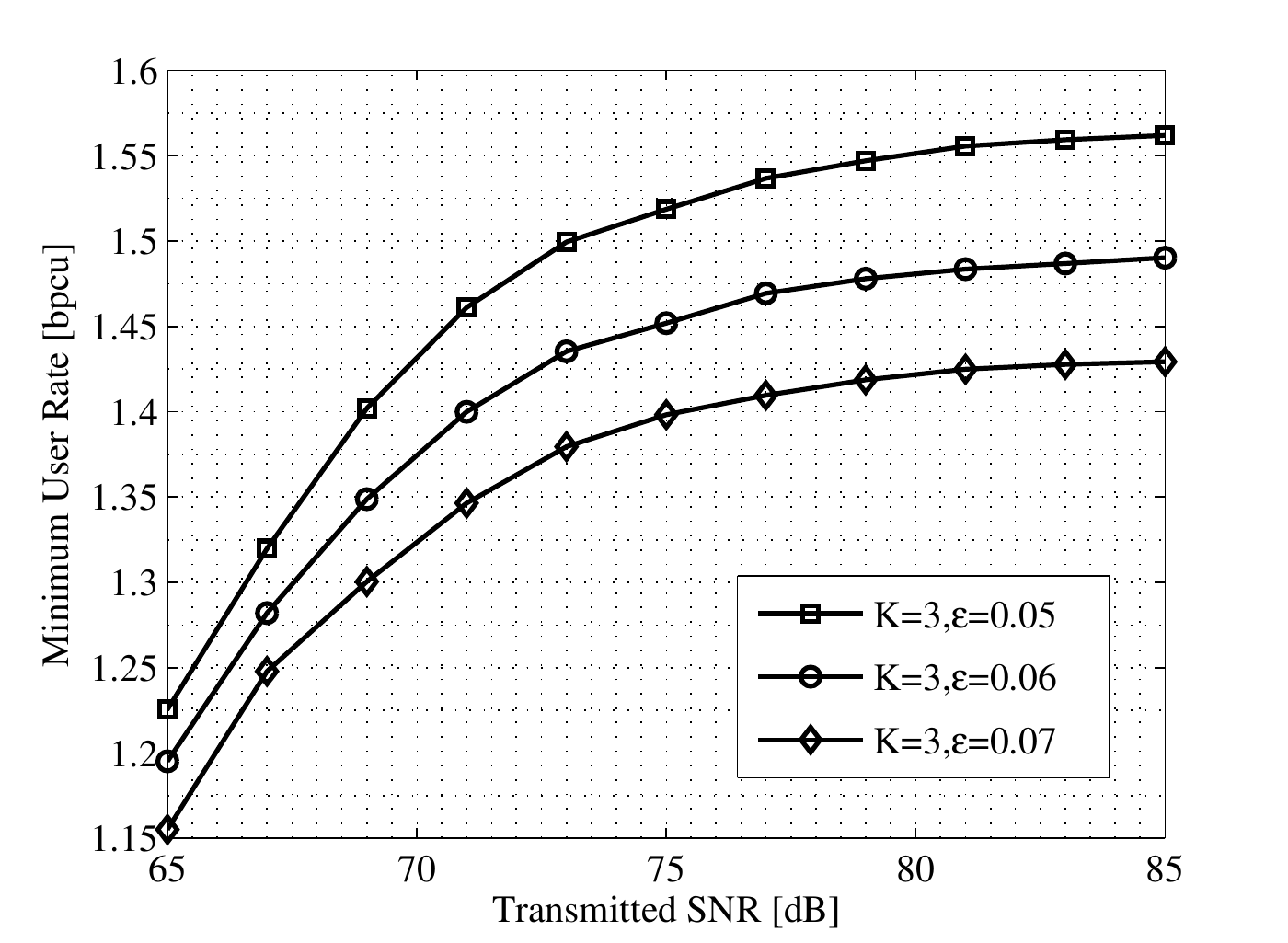}
\caption{The maximized minimum user rate comparison under different residual interference coefficient $\varepsilon$ values when $\bm{T}=[0.6,0.6,0.6]$.}
\label{R_min_max_SNR}
\end{figure}

According to the QoS-guaranteed max-min user rate criterion, the maximized minimum user rate is achieved at $\bm{R_1}=[1.427,1.427,1.456]$ for $K=3$, $TSNR=70\text{dB}$, $\varepsilon=0.05$, $\bm{H}=[0.293,0.359,0.454] \times 10^{-4}$ and $\bm{T_1}=[1,1,1]$. When one of the users requests higher rate as $\bm{T_2}=[2,1,1]$, the solution turns into $\bm{R_2}=[2.000,1.158,1.160]$. Obviously, the minimum user rate changes from $1.427$ to $1.158$ under the extra QoS constraints. Similar to the max-sum rate criterion aforementioned, the residual interference can reduce the maximized minimum rate, as showed in Figure \ref{R_min_max_SNR}.


\section{Conclusion}

In this letter, we propose the multiple access and interference management scheme for NOMA VLC network. With residual interference during SIC taken into account, we investigate the QoS-guaranteed power allocation for NOMA based on either max-sum rate criterion or max-min rate criterion. By virtue of the gradient projection algorithm, the power allocation coefficients can be dynamically adjusted.





\bibliographystyle{IEEEtran}
\bibliography{paper}
%

\end{document}